\magnification=1200
\baselineskip=18truept

\def\chiral{{\rm{\bf C}}}
\def\wilson#1{{\rm{\bf B^#1}}}
\def\ham#1{{\rm{\bf H^#1}}}

\line{\hfill UW/PT--96--02}
\line{\hfill DOE/ER/40561-246-INT96-00-115}
\line{\hfill RU--96--9}
\vskip 2truecm
\centerline{\bf Overlap formulation of Majorana--Weyl fermions.}

\vskip 1truecm
\centerline{Patrick Huet$^{1}$, Rajamani
Narayanan$^{2}$ and Herbert Neuberger$^{3}$}
\vskip 0.5truecm
\centerline{${}^1$ \it Department of Physics, Box 351560}
\centerline{\it University of Washington, Seattle, WA 98195-1560.}
\vskip .1in
\centerline{${}^2$ \it Institute for Nuclear Theory, Box 351550}
\centerline{\it University of Washington, Seattle, WA 98195-1550.}
\vskip .1in
\centerline{${}^3$\it Department of Physics and Astronomy}
\centerline{\it Rutgers University, Piscataway, NJ 08855-0849. }
\vfill
\centerline{\bf Abstract}
\vskip 0.75truecm
An overlap method for regularizing Majorana--Weyl fermions interacting
with gauge fields is presented. A mod(2) index is introduced in
relation to the anomalous violation of a discrete global chiral
symmetry. Most of the paper is restricted to 2 dimensions but
generalizations to 2+8k dimensions should be straightforward.
\vfill
\eject
In 2+8k dimensional Minkowski space the most basic fermion is the
Majorana--Weyl (MW) fermion. MW fermions, when in
irreducible real representations
of Lie groups, can interact with the relevant gauge bosons. There exist
gauge theories that cannot be viewed as containing only Weyl
fermions. This is not true in four dimensions, and, for
this reason, the overlap formalism was originally
set up to deal only with the Weyl case [1]. In this paper we extend
this formulation to MW fermions. For simplicity we work in 2 dimensions,
but we believe that the main points generalize to other
appropriate dimensions. A case of particular interest to us
is ten dimensional N=1 supersymmetric Yang Mills; we hope to use a
regulated version of this (non-renormalizable, anomalous) gauge
theory as a means to regulate the four dimensional N=4
supersymmetric Yang Mills theory. The
latter is interesting for many reasons, in particular, the large
$n$ limit of the $SU(n)$ case would have a cutoff independent
critical value of the coupling constant where arbitrarily large
planar Feynman diagrams dominate [2].

We work in Euclidean space and physical charge conjugation does not have
a natural analogue [3]. What we mean then by MW is that a system of
Weyl fermions in a real representation interacting with an external
gauge field decouples into two isomorphic sub--systems and each such
sub--system describes a set of MW fermions. Let the Weyl system be
described by the action,
$$
{\cal L} =\int {\bar\psi_L} (\sigma \cdot (\partial + i A ))\psi_L
\equiv \int \bar\psi_L C \psi_L ,$$
with $A_\mu =-A^*_\mu =A^\dagger_\mu
$ and $\sigma_\mu = (1,i)$ for $\mu = (1,2)$.
${\bar\psi_L}$ and $\psi_L$ are independent Grassmann variables.
All group indices are suppressed.
The purely imaginary character of
$A_\mu$ (hermiticity is always assumed) is equivalent
to a skew-symmetry of $C$: For any two ordinary complex
functions $\phi_1$ and
$\phi_2$, $\int \phi_1 C \phi_2 = -\int \phi_2 C \phi_1 $.  This
skewness leads to the above mentioned decoupling and, consequently
to a factorization of the determinant of $C$. Defining
$\bar\psi_L=(\xi+i\eta)/\sqrt{2}$ and $\psi_L=(\xi-i\eta)/\sqrt{2}$
with $\xi$ and $\eta$ independent (real) Grassmann variables we
find:
$$
{\cal L}={1\over 2} \int \xi C \xi + {1\over 2} \int \eta C \eta .$$
For fixed $A_\mu$, correlation functions of the $\psi$'s are
given in a fixed pattern of sums of
products of correlation functions of the $\xi$'s and the $\eta$'s.
By analytic continuation to Minkowski space the standard expressions
are obtained.

The Weyl theory has a $U(1)$ global (chiral) symmetry, potentially violated by
quantum effects when the $A_\mu$-fields are made dynamical. In the
MW basis the $U(1)$ appears as an $O(2)$ rotating $\xi$ and $\eta$
into each other. If we go to a single MW system the $O(2)$ disappears and
all we are left with is a discrete $Z_2$ flipping the sign of $\xi$.
In an anomaly free gauge theory one has to have both left--handed
and right--handed fields and there will be several such $Z_2$'s
(some combinations of these $Z_2$'s
are not chiral). The chiral $Z_2$'s forbid mass terms, and
would imply the absence of bilinear fermionic condensates in finite
Euclidean physical volumes.
(The symmetries are discrete, so breaking them
spontaneously in the infinite volume limit, even in two dimensions,
is possible.)

However, the ancestry of these symmetries at the Weyl level
indicates that, under certain circumstances,
one should expect explicit violations of the
global axial symmetries. For Weyl
fermions the explicit breaking is caused by topologically
nontrivial gauge backgrounds. The associated fermionic zero modes
lead to non-vanishing symmetry-breaking condensates [4]. The
stable part of the number of zero modes comes from certain
$C$'s that have a nonzero analytical index
($dim(Ker( C)) - dim (Ker( C^\dagger))$). Not for every type of gauge field
does the possibility of a non-zero index even arise, but
the option exists generically, and some gauge fields realize it.
The above is well known.

In the MW case $C^\dagger = - C^*$ and the analytical index
vanishes. One can still define a modulus 2 index in this case:
The parity of the dimension of the kernel of $C$ is invariant
under small deformations of $C$ subjected to skewness. We sketch a
physicist's proof of this below.

We are on a compact manifold and $A_\mu$ is bounded; therefore
$dim(Ker (C))$ is guaranteed to be a finite number, say $k$.
We are interested in how $k$ would change under a
perturbation of $C$. Let the
space $C$ acts on be denoted by $V$. On $V \oplus V$ define
$D=\pmatrix{0&C\cr -C^* & 0 } = D^\dagger$. If
$Ker (C )= span ~[ u_1, u_2,...,u_k ]$, then, the kernel of $D$
is given by:
$$span ~
\biggl[ \pmatrix{0 \cr u_1}, \pmatrix{0 \cr u_2},...,
\pmatrix{0\cr u_k};\pmatrix{ u_1^* \cr 0}, \pmatrix{u_2^* \cr
0},...,\pmatrix{u_k^* \cr 0 }\biggr]\equiv
span ~[ \psi_{\alpha},~\alpha=1,2...,2k ].$$
Since $D$ is hermitian we apply ordinary
degenerate perturbation theory. We should diagonalize $O$,
the perturbation matrix of $D$ restricted to the kernel of $D$,
$$
O_{\alpha,\beta} = \int \psi^\dagger_\alpha \delta D \psi_\beta .$$
The perturbation in $D$ was
induced by a perturbation in $C$, so $O$ has the structure
$$
O=\pmatrix{0&A\cr -A^* & 0} ,$$
with $A^t = -A;~A_{ij} = \int u_i \delta C u_j$. Since
the matrix $A$ is antisymmetric its rank must be even. Hence,
the perturbation will not be able to change the parity of $k$.

If the mod(2) index is odd ($(-)^k =-1$) there will be at least one
zero mode stable under small deformations of the background gauge field
and a bilinear condensate is possible, explicitly violating
some of the chiral $Z_2$'s. Conversely, an odd mod(2) index is
a necessary condition for a non-vanishing bilinear expectation value
in finite Euclidean volumes.

We now turn to the overlap formalism. Our objective is to find
the analogous decoupling of a system of Weyl fermions in a real
representation into MW fermions and understand how the discrete
version of anomalous symmetry is realized.
Any reasonable regularization of chiral
gauge theories which preserves the bilinearity
of the action like the overlap does,
must provide equivalent realizations of decoupling and anomalous
discrete symmetries.

We start from the Weyl overlap. The fermion induced action
corresponding to a left--handed  Weyl multiplet
$
z_W [A_\mu ] = \int [d\bar\psi_L d\psi_L ] e^{-{\cal L}}$
is replaced on the lattice by the overlap
$
z^{\rm lat}_W [U_\mu ] = {}_U\!\! <L-|L+>_{U}$.
The lattice is a symmetric torus of linear size $l$ with $l$ even.
This ensures that the total number of MW degrees of freedom
is even.
The lattice link variables $U_\mu$ replace the continuum $A_\mu$'s
and, in addition, implement the fermionic boundary conditions.
The states $|L\pm>_U$ are the ground states of two many body hamiltonians
$
{\cal H}^\pm =\sum_{x\alpha,y\beta} a^\dagger_{x,\alpha }
\ham\pm(x\alpha,y\beta; U)
a_{y,\beta },~~
\{a^\dagger_{x,\alpha }, a_{y,\beta }\} =\delta_{\alpha,\beta}\delta_{x,y},$
$\alpha,\beta=1,2$ and $x=(x_1 , x_2 )$ with
$x_\mu =0,1,2,.....,l-1 $. The single particle hermitian
hamiltonians $\ham\pm$ are given by:
$$\eqalign{
\ham\pm & =\pmatrix {\wilson\pm &\chiral\cr \chiral^\dagger&-\wilson\pm},\cr
\chiral(x,y) & ={1\over 2}
\sum_\mu \sigma_\mu (\delta_{y,x+\mu}U_\mu (x) ~- ~
\delta_{x,y+\mu}U_\mu^t (y)),\cr
\wilson\pm (x,y) & = {1\over 2} \sum_\mu
(2\delta_{x,y}~ - ~\delta_{y,x+\mu}U_\mu (x)~ - ~
\delta_{x,y+\mu}U_\mu^t (y))~\pm~ m\delta_{x,y} .\cr}
$$
The parameter $m$ is restricted only by $0 < m < 2$. Group indices have been
suppressed. The phases of the states $|L\pm>_U$ are chosen according to the
Wigner-Brillouin convention [1]; they are irrelevant for the subsequent
analysis.

First we wish to decouple the systems to attain a factorization of the overlap.
Suppressing also the site indices we define a new set of creation/annihilation
operators: $a_1 = {{\xi - i\eta }\over\sqrt{2}}, a_2 =
{{\xi^\dagger - i\eta^\dagger }\over\sqrt{2}}$. The transformation
is canonical. Under a Euclidean space rotation, both $\xi$ and $\eta$
transform as left--handed fields. Under a $U(1)$ phase transformation
of the $a_\alpha$'s, $\xi$ and $\eta$ undergo an $O(2)$ transformation.
The explicit form above ensures that the matrices $\wilson\pm$ are real.
A short computation shows that
$$
{\cal H}^\pm = {1\over 2} \alpha^\dagger \ham\pm \alpha +
{1\over 2} \beta^\dagger \ham\pm \beta ,$$
with all indices suppressed, $\alpha_1 = \xi , \alpha_2 = \xi^\dagger$
and $\beta_1 = \eta , \beta_2 = \eta^\dagger$. The two terms
above commute with each other, are isomorphic to each other
and act in different spaces;
thus the overlap will factorize into two equal factors.

We now turn to the mod(2) index.
How would we solve for the overlap if we had a single MW multiplet
and did not want to use the fact that it is the square root of
a Weyl system ?
To answer this question it is
useful to change bases again: Define two hermitian operators,
$\gamma_1 = {{\xi+\xi^\dagger}\over\sqrt{2}}$ and
$\gamma_2 = {{\xi-\xi^\dagger}\over{i\sqrt{2}}}$. They obey the
canonical commutation relations
$\{\gamma_\alpha ,\gamma_\beta \} =\delta_{\alpha, \beta}$. The collection
of all the $\gamma$'s, with all indices taken into account, generates
a large Clifford algebra. Our hamiltonians are bilinears in this
algebra given by
$$
{\cal H}^\pm_{mw} = {1\over 2} \gamma \ham\pm_{mw} \gamma$$
with all indices suppressed.  The hamiltonian matrices are purely
imaginary and antisymmetric and take the form
$$
\ham\pm_{mw} = \Gamma_{3} \left( \Gamma_{1} {\rm Re} \chiral +
                               \Gamma_{2} {\rm Im} \chiral +
                               \wilson\pm \right),
$$
where we have introduced real symmetric two dimensional Dirac
$\Gamma$-matrices, $\Gamma_1 = \sigma_3, \Gamma_2=\sigma_1$ and picked
$\Gamma_3 =-\sigma_2$, which is antisymmetric; the $\sigma_i , \,
i=1,2,3$ are the Pauli matrices.  The $\ham\pm_{mw}$ are recognized as
$\Gamma_3$ times lattice Wilson-Dirac operators, one with a positive
mass term and the other with a negative one.  This structure will
generalize to other dimensions where MW fermions exist.

We need to carry out Bogolyubov transformations to bring the
hamiltonians to some canonical form. This amounts to an orthogonal
transformation which obviously preserves the anticommutation
relations among the $\gamma$'s. The $\ham\pm_{mw}$ can be brought to a
canonical form
where all nonzero elements are restricted to antisymmetric two by two
blocks along the diagonal.
One can arrange the left bottom entry of each block to
be a positive
number $\lambda$ times $i$
and order the blocks by the magnitude of $\lambda$
(generically, there will be no degeneracies).
One such two by two block would correspond to
a term $i\lambda(\gamma_2 \gamma_1 - \gamma_1 \gamma_2 )$. Here
the $\gamma$'s are the new, rotated, canonical generators. Rewriting the
$\gamma$'s in terms of new $\xi$'s this term becomes
$\lambda (\xi\xi^\dagger - \xi^\dagger \xi )$ indicating that in
the ground state the appropriate $\xi$ single particle state should
be filled.

The Hilbert space splits into ``even'' and
``odd'' subspaces; on the even
subspace the product of all the $\gamma$'s in some fixed order
(the ``chirality'') is $\zeta$,
while on the odd subspace this product is $-\zeta$ ($\zeta =\pm 1$ or
=$\pm i$).
The hamiltonians do not connect these two
subspaces. There is then the possibility that one of the
$|L \pm >_U $ ground states
be odd and the other even. The way to ascertain whether this
happens or not, is to look at the combined orthogonal transformations
which, individually, bring each hamiltonian matrix to its
canonical form. The combined orthogonal matrix has
a determinant equal to $\pm 1$. If it is $1$ the parities of the
two ground states are the same; if it is $-1$ they are opposite. When
they are opposite the overlap will vanish; in the continuum the
vanishing would be attributed to $C$ having an odd number of zero modes.
If $C$ has a single zero mode the expectation value of one fermion field
could be non-zero. Similarly, in the lattice overlap formulation,
the insertion of a single $\gamma$ between the states will restore
equal parity and will, generically, lead to a non-vanishing result.

Let us now explain why
the sign of the determinant of the big orthogonal transformation,
$O$,
connecting the two bases in which the individual hamiltonians
have their canonical forms
indeed is the mod(2) index in the overlap formulation. Assume
first that $\det (O) = 1$. Then, there exists an antisymmetric real
matrix $T$ such that $O=e^T$. The unitary operator
${\cal U}$ which implements the canonical transformation
$\gamma^\prime = O \gamma = {\cal U}^\dagger \gamma {\cal U} $
is given by ${\cal U} = e^{{1\over 2}\gamma T \gamma}$. If
$|L->_{U}$ is one of the ground states (why we pick the minus sign
will become clear in the next
paragraph, but is immaterial for the present discussion)
the other is $|L+>_U = {\cal U}^\dagger |L->_U$. It is now
made very explicit that the two states $|L\pm >_U$ have identical
parities. Assume now that $\det (O) = -1$. Define
a canonical transformation $\gamma^\prime =
\gamma_{\sharp} \gamma \gamma_{\sharp} = O_1 \gamma$
implemented by one particular $\gamma_{\sharp}$.
The matrix $O_1$ is diagonal and has all entries $-1$
except the one associated with the chosen $\gamma_{\sharp}$ which is $+1$. It
represents a ``parity'' transformation switching
the odd and even ``chiralities''.
Since the
total space is even dimensional, $\det (O_1 ) =-1$. The
unitary operator ${\cal U}$ implementing the
canonical transformation defined by $O O_1$ is worked
out just as before since $\det (O O_1)=1$. The unitary
operator implementing the original canonical transformation
$O$ is now ${\cal U}\gamma_{\sharp}$ and, just as above, we now arrive at
the conclusion that the two ground states have opposite parities.
In summary,
the continuum $(-1)^{dim(Ker(C))}$ corresponds on the lattice to $\det(O)$.
Note that on the lattice $(-1)^{dim(Ker(\chiral))}=1$ always, showing
that it would be mistaken to replace
the continuum $C$ by a finite matrix of rigid form.

For ${\cal H}^-_{mw}$ the ground state has the same parity for all
gauge fields; the proof of this goes as follows.  For very large
values of $|m|$ the state is obviously independent of the gauge
fields.  Let us increase $m$ towards some finite negative value. If
the parity is to change at some mass value, some filled state must
become empty at that value, at which point the associated $\lambda$
vanishes.  But we know already from the Weyl case that $\ham-$ always
has a gap [1]. Therefore, as long as the mass is negative, the parity
cannot change. We can choose the parity operator such that the ground
state of ${\cal H}^-_{mw}$ is even.

It remains to be shown that there are
actual instances in which
${\cal H}^+_{mw}$ has an odd ground state. This is
trivial, since
we could view the ordinary $U(1)$ Schwinger model
[5] as an $SO(2)$
gauge theory and each Weyl fermion as a doublet
of Majorana ones of the same handedness. The instantons of $U(1)$,
when viewed as $SO(2)$ configurations, make the required ground state odd. Of
course, this is a very special case where
MW fermions aren't necessary and there exists a more discerning
index. However,
embedding the $SO(2)$ into any $SO(n)$ in a trivial way
shows that whenever the gauge
group is $SO(n)$ and the fermions are in the vector
(defining) representation
of $SO(n)$ odd vacua will appear.
Since this is the most general case
the appearance of odd states is not an isolated event.

For odd states to appear in a way that
has a chance to survive in the continuum limit
it is necessary that the space of
gauge fields in the continuum be disconnected. In two dimensions
this will happen
if the ``true'' gauge group, i.e. the one that is seen by the fermions, is
multiply connected. Then this gauge group
can be written as a simply connected (covering) group
divided out by a nontrivial subgroup of its center. The center is always
a subgroup of $Z$.
Clearly, the ``true'' group for
the fermions in the vector representation of $SO(n)$,
$SO(n)=Spin(n)/Z(2)$, is of this type.

Another interesting example is provided by
``adjoint QCD'', where the
``true'' gauge group is $SU(n)/Z(n)$.
(For $n=2$, this is just the $SO(3)$ case
above.) Let the Cartan generator
given by ${1\over\sqrt{n(n-1)}}
diag (1,1,1,...,1,1-n)$ be denoted by $H$, normalized by $tr(H^2) =1$.
Embedding a $U(1)$ instanton in $SU(n)/Z(n)$ in the $H$ direction
yields $n-1$ zero modes for $C$ [6];
these zero modes provide an $n-1$ dimensional
representation of the discrete symmetry
of $H$ which permutes the first $n-1$ entries.  We conclude that, for
$n$ even, odd ground states $|L+>_U$ will occur.
A more thorough examination of this
case is reserved for the future. The two immediate issues
to be resolved are whether indeed there is a fundamental difference
between odd and even $n$'s [7]
(this is important also for the large $n$ limit),
and
whether for even $n$'s larger than 2 one has non-vanishing bilinear
condensates in finite Euclidean volumes. The overlap formulation
is guaranteed to provide a numerical tool well suited to this
problem and complementary to analytical methods. No other numerical
approach we are aware of could be of similar use.
We plan to address similar issues regarding
gluino bilinear condensates in four
dimensional N=1 pure supersymmetric Yang Mills theories.

\vskip .2cm

\noindent {\bf Acknowledgments}

This research was supported in part by the DOE under grants
\# DE-FG06-91ER40614 (PH and RN),
\# DE-FG06-90ER40561 (RN) and
\# DE-FG05-90ER40559 (HN).
\vfill\eject

\noindent{\bf References}
\vskip .2in
\item{[1]}  R. Narayanan, H. Neuberger, Phys. Lett. B302 (1993) 62;
Nucl. Phys. B412 (1994) 574; Nucl. Phys. B443 (1995) 305.
S. Randjbar--Daemi, J. Strathdee, Phys. Rev. D51 (1995) 6617;
Nucl. Phys. B443 (1995) 386; Phys. Lett. B348 (1995) 543;
hep-th \# 9510067 (IC/95/305).
\item{[2]} H. Neuberger, Nucl. Phys. B. (Proc. Suppl.){\bf 26} (1992) 563.
\item{[3]} P. Ramond, {\it Field Theory, A Modern Primer} (Benjamin/Cummings,
Reading, MA, 1981).
\item{[4]} G. 't Hooft, Phys. Rev. Lett. 37 (1976) 8; Phys. Rev. D14 (1976)
3432.
\item{[5]} I. Sachs, A. Wipf, Helv. Phys. Acta 65 (1992) 652;
R. Narayanan, H. Neuberger, P. Vranas, Phys. Lett. B353 (1995) 507.
\item{[6]} A. Smilga, Phys. Rev. D49 (1994) 6836.
\item{[7]} S. Lenz, M. Shifman, M. Thies, Phys. Rev. D51 (1995) 7060.
\vfill
\eject

\end